\documentclass[aps,pra,reprint,amsmath,amssymb,superscriptaddress,floatfix]{revtex4-2}

\usepackage{amsmath,amssymb,mathtools,amsthm,bm}
\usepackage{booktabs,tabularx,array,multirow}
\usepackage{graphicx}

\usepackage{enumitem}
\usepackage{microtype}
\usepackage{hyperref}

\allowdisplaybreaks[4] 

\usepackage{xcolor}

\hypersetup{
  colorlinks=true,
  linkcolor=blue, 
  citecolor=blue,  
  urlcolor=black,
  pdfauthor={Haihui Wang},
  pdftitle={Efficient Pauli Truncation from Operator Stabilizer R\'enyi Entropy}
}

\setlist[itemize]{leftmargin=1.6em,itemsep=0.2em,topsep=0.3em}
\setlist[enumerate]{leftmargin=1.8em,itemsep=0.2em,topsep=0.3em}
\renewcommand{\arraystretch}{1.18}

\theoremstyle{plain}
\newtheorem{theorem}{Theorem}
\newtheorem{lemma}{Lemma}

\theoremstyle{definition}
\newtheorem{definition}{Definition}

\newtheorem{remark}{Remark}

\newcommand{\PN}{\mathcal P_N}
\newcommand{\E}{\mathbb E}
\newcommand{\tr}{\operatorname{tr}}
\newcommand{\Top}{\operatorname{Top}}

\newcommand{\OTOC}{\operatorname{OTOC}}
\newcommand{\M}{M}

\newcolumntype{Y}{>{\raggedright\arraybackslash}X}

\ifdefined\pdfminorversion
\fi
\begin{document}
\title{Efficient Pauli Truncation from Operator Stabilizer R\'enyi Entropy}
\raggedbottom

\author{Haihui Wang}

\author{Yongming Li}
\email[Contact author: ]{liyongm@snnu.edu.cn}

\affiliation{School of Mathematics and Statistics,
Shaanxi Normal University, Xi'an 710119, China}

\begin{abstract}
Operator stabilizer R\'enyi entropy (OSE) quantifies the magic of Heisenberg operators and serves as a diagnostic of their compressibility in the Pauli basis.
We establish rigorous bounds showing when an operator can be efficiently approximated by Pauli truncation in terms of the scaling of its OSE.
Specifically, we prove that volume-law scaling of OSE for $\alpha\geq1$ prevents efficient Pauli truncation from reproducing expectation values for all states at sufficiently small fixed error. On the other hand, logarithmic growth of OSE for $0<\alpha<1/2$ guarantees efficient Pauli truncation approximability uniformly over all states.
A counterexample shows that this sufficient range cannot be extended on the basis of OSE scaling alone.
If we restrict to correlations over suitable low-average state ensembles, then the sufficient range extends to $0<\alpha<1$.
We further derive cumulative error bounds for iterative Pauli propagation and approximation guarantees for fixed-order out-of-time-ordered correlators.
Moreover, under an independent-sign assumption, we show that logarithmic OSE for $0<\alpha<1$ guarantees efficient Pauli truncation with the prescribed accuracy for all input states with high probability.
Our results provide a theoretical framework for understanding when operator magic constrains or enables efficient Pauli truncation and for guiding the development of classical
simulation methods.

\end{abstract}
\maketitle

\section{Introduction}
Classical simulation of quantum dynamics is central to understanding non-equilibrium many-body physics and assessing the capabilities of quantum computers~\cite{Feynman1982}.
Accurate numerical methods for evolving quantum states or observables in time provide access to non-equilibrium phenomena, including thermalization ~\cite{Deutsch1991,Srednicki1994,DAlessio2016}, transport~\cite{Bertini2021Transport}, and information scrambling~\cite{Nahum2018,vonKeyserlingk2018}.
However, the exponential growth of Hilbert space makes direct state evolution costly.
Exact diagonalization is limited to small systems by memory requirements.
Tensor-network approaches, closely related to the density-matrix
renormalization group, provide an important alternative~\cite{White1992,Vidal2004,Haegeman2011,Schollwock2011,Orus2014,Cirac2021}.
Their effectiveness is closely related to entanglement structure and
area-law behavior~\cite{Vidal2003,Eisert2010}, while entanglement growth
can restrict accessible simulation times~\cite{CalabreseCardy2005}.
To address the  limitations  of traditional approaches, some studies have shifted perspective to the Heisenberg picture~\cite{Prosen2007,Hartmann2009}, focusing on the evolution of operators rather than states. Although physically equivalent to state evolution, evolving the relevant observables in the Heisenberg picture can offer an exponential computational advantage in suitable settings, enabling calculations of transport and scrambling without explicitly representing the full time-evolved state~\cite{RakovszkyEtAl2022,XuSwingle2020}.
Correlation functions provide physically relevant targets for this approach: infinite-temperature correlations characterize transport and relaxation~\cite{Bertini2021Transport,RakovszkyEtAl2022}, while out-of-time-ordered correlators (OTOCs) probe operator spreading and information scrambling~\cite{Nahum2018,vonKeyserlingk2018,XuSwingle2020}. Infinite-temperature correlations and averages over computational-basis states also motivate approximation guarantees for ensembles whose average state is sufficiently mixed~\cite{Dowling2026LOE,Schuster2025Polynomial}.

The benefit of observable evolution depends on whether the evolving operators admit compact representations.
Pauli propagation provides a natural approach for qubit systems by expanding an observable in Pauli strings and updating their coefficients under the dynamics~\cite{Rall2019,RudolphEtAl2025}.
Clifford conjugation maps each Pauli string to a single string, whereas non-Clifford Pauli rotations can generate additional terms.
The resulting growth raises a central question: which intrinsic properties of an operator permit accurate approximations with a manageable number of Pauli terms?
Pauli truncation reduces the number of terms in an operator's Pauli expansion by discarding selected Pauli terms. Different truncation strategies use criteria such as coefficient magnitude, Pauli weight, or Pauli-path weight~\cite{RudolphEtAl2025}.
The methods based on Pauli truncation have been applied to quantum-circuit simulation and many-body dynamics in two and three dimensions~\cite{BegusicHejaziChan2025,BegusicChan2025}. For noisy circuits~\cite{ShaoWeiChengLiu2024,Schuster2025Polynomial,AngrisaniMele2026}, suitable ensembles of noiseless circuits~\cite{AngrisaniEtAl2025}, and short-time Hamiltonian dynamics with sufficiently entangled input states~\cite{XuEtAl2026}, provable error bounds have been established. These advances motivate quantitative criteria relating operator structure to the number of retained Pauli
terms and the resulting approximation error.

The operator stabilizer R\'enyi entropy (OSE) quantifies operator magic through the distribution of squared Pauli coefficients and has an operational interpretation in
terms of approximation by a small number of Pauli strings~\cite{DowlingKosTurkeshi2025}.
Its resource-theoretic motivation is part of a broader effort to connect stabilizer structure~\cite{Leone2022,WangLi2023Qudits}
 and magic with classical simulation~\cite{AaronsonGottesman2004,BravyiSmithSmolin2016,HowardCampbell2017}.
Related developments include channel Pauli spectrum, magic-channel distillation, and probabilistic channel simulation~\cite{WangLi2025Channels,WangLi2026Simulation}.
The scaling of OSE was related to the spectral norm truncation error, which bounds expectation value errors for all states~\cite{DowlingKosTurkeshi2025}. Specifically, an extensive 1-R\'enyi OSE requires exponentially many retained Pauli strings to achieve a sufficiently small fixed error, ruling out efficient simulation by Pauli truncation,  whereas logarithmic 0-R\'enyi OSE guarantees an exact Pauli representation with only polynomially many terms.
Subsequent work derived OSE-dependent bounds on the squared-coefficient tail, yielding explicit truncation budgets for a prescribed normalized Hilbert--Schmidt error, with applications to quantum spin dynamics~\cite{ShaoChengLiu2026}. However, normalized Hilbert--Schmidt and spectral norm errors correspond to different accuracy requirements and need not exhibit the same scaling~\cite{Bhatia1997}.
This distinction motivates determining which R\'enyi indices allow logarithmic OSE growth to guarantee accurate expectation values
over all input states.
Further developments have bridged OSE and local-operator entanglement (LOE)~\cite{DowlingModiWhite2025} and used it to characterize Pauli spectrum and noise-induced simulability transitions in random circuits~\cite{DowlingEtAl2026Noise}.
Therefore, OSE provides a useful indicator for assessing the classical simulability of quantum operators through Pauli truncation, in a way analogous to how entanglement entropy characterizes the efficiency of tensor network methods.

A natural question is how OSE scaling characterizes the efficiency of Pauli truncation. While the cases $\alpha=0$ and $\alpha=1$ have been studied~\cite{DowlingKosTurkeshi2025},
the expectation value guarantees for all states at other R\'enyi indices  require further investigation.
Can OSE at different R\'enyi indices provide necessary lower bounds on the number of Pauli terms required to achieve a prescribed approximation accuracy? Conversely, for which R\'enyi indices does logarithmic OSE growth guarantee that polynomially many retained Pauli terms suffice to achieve the prescribed accuracy uniformly over all input states?

In this paper, we answer those questions.
We establish theoretical guarantees for Pauli truncation of Heisenberg operators by relating OSE scaling to the number of Pauli terms required to achieve a prescribed accuracy.
The Pauli truncation discussed here refers to the strategy of retaining the $\chi$ largest magnitude Pauli coefficients, which is known as Top-$\chi$ Pauli truncation~\cite{DowlingKosTurkeshi2025,ShaoChengLiu2026}.
We first prove that linear growth of OSE at $\alpha=1$, or power-law growth at $\alpha>1$ necessarily implies that Pauli truncation cannot efficiently reproduce expectation values for all states. Conversely, we show that logarithmic growth of OSE for  $0<\alpha<1/2$ implies efficient Pauli truncation approximability for all states.
We give a counterexample for $\alpha\geq1/2$ that establishes the sharpness of this sufficient range. For correlations averaged over low-average state ensembles with a polynomially bounded parameter, the sufficient range extends to $0<\alpha<1$. We further control cumulative errors in iterative Pauli propagation and extend the approximation guarantees to higher-order OTOCs of fixed order.
Finally, under an additional independent-sign assumption, we obtain spectral norm error bounds in expectation and with high probability for $0<\alpha<1$. Finite-system numerical results illustrate the analytical bounds and the distinctions between the different accuracy requirements.

The rest of this paper is organized as follows. Section~\ref{sec:setup} introduces several useful results on OSE and Pauli truncation. Section~\ref{sec:truncation-guarantees} establishes the necessary bounds and sufficient conditions for Pauli truncation. Section~\ref{sec:iterative-pauli} develops cumulative error control for iterative propagation and discusses computational cost and readout. Section~\ref{sec:otoc} extends the analysis to higher-order correlators, and Section~\ref{sec:analysis-error} studies spectral truncation error under independent random signs. Section~\ref{sec:conclusion} provides conclusions and future directions.
\section{Operator stabilizer R\'enyi entropy and Pauli truncation}
\label{sec:setup}
In this section we will introduce the setting and notation, and review some useful results.
We denote by $\PN=\{I,X,Y,Z\}^{\otimes N}$ the set of Pauli strings acting on the $N$-qubit Hilbert space $\mathcal{H}=(\mathbb{C}^2)^{\otimes N}$, whose total dimension is $D=2^N$.
Unless explicitly stated otherwise, $O$ is a Hermitian operator satisfying the Hilbert--Schmidt normalization  $D^{-1}\tr(O^2)=1$.
We use base-$2$ logarithms, denoted by $\log$, and $\ln$ denotes the natural logarithm.
The Schatten norms are $\|A\|_p=(\tr|A|^p)^{1/p}$.
In particular, $\|A\|_2$ is the Hilbert--Schmidt norm, and $\|A\|_\infty$ is the spectral norm~\cite{Bhatia1997}.
The notations $f(N)=\mathcal{O}(g(N))$ and
$f(N)=\Omega(g(N))$ mean that
$0\leq f(N)\leq Cg(N)$ and $f(N)\geq cg(N)\geq 0$
for all sufficiently large $N$, respectively,
where $C,c>0$ are constants independent of $N$.

In the Heisenberg picture, we consider the ordered Pauli decomposition of the evolving operator
\begin{equation}
\label{eq:decomposition}
\begin{aligned}
O_N&=U_t^\dagger O U_t=\sum_{i=1}^r a_i P_i,\\
a_i&=D^{-1}\tr(P_iO_N),\quad |a_1|\geq\cdots\geq|a_r|>0,
\end{aligned}
\end{equation}
where $U_t$ is a unitary evolution and $P_i\in\PN$ are distinct Pauli strings satisfying $D^{-1}\tr(P_iP_j)=\delta_{ij}$.
Since unitary conjugation is Hermitian-preserving and preserves the Hilbert--Schmidt norm, $O_N$ is Hermitian and satisfies $D^{-1}\tr(O_N^2)=1$.
Consequently, $a_i\in\mathbb R$ and $\sum_{i=1}^{r}|a_i|^2=1$.
The operator stabilizer R\'enyi entropy (OSE) of $O_N$ is defined as~\cite{DowlingKosTurkeshi2025}
\[\mathcal{M}^{(\alpha)}(O_N) :
=\frac{1}{1 - \alpha}\log\sum_{i=1}^{r}|a_i|^{2\alpha},\quad
 \alpha>0,\ \alpha\neq1.\]
The limiting cases  are $\M^{(0)}=\log r$, $\M^{(1)}=-\sum_i|a_i|^2\log|a_i|^2$, and $\M^{(\infty)}=-\log|a_1|^2$.

A fundamental problem is to compute expectation values of observables
\begin{equation}
\label{eq:expectation-value}
\langle O\rangle_t
=\operatorname{tr}\!\left[O\,U_t\rho U_t^\dagger\right]
=\operatorname{tr}\!\left[O_N\rho\right],
\end{equation}
where $O$ is an observable and $\rho$ is an arbitrary initial state.
Direct evaluation of Eq.~\eqref{eq:expectation-value} based on the Pauli decomposition of $O_N$ in Eq.~\eqref{eq:decomposition}  involves a sum over $r$ Pauli terms. However, $r$ can grow exponentially with the number of qubits,  increasing both computational and storage requirements~\cite{RudolphEtAl2025}.

To reduce these costs, one truncation strategy for approximating the operator $O_N$ consists of retaining only the largest $\chi$ amplitudes $|a_i|$ in Eq.~\eqref{eq:decomposition}, which is called Top-$\chi$ truncation~\cite{DowlingKosTurkeshi2025,ShaoChengLiu2026}.
For an integer $\chi$ with $0\leq \chi \leq r$, we define the truncated operator
$\widetilde O_{N,\chi}=\Top_\chi(O_N) :=\sum_{i=1}^{\chi}a_iP_i$.
When $\chi=0$, the truncated operator vanishes, $\widetilde{O}_N = 0$; when $\chi = r$, no truncation occurs and $\widetilde{O}_N = O_N$.
Unless otherwise specified, we only consider the nontrivial case  $0< \chi < r$.
We write $\Delta O=O_N-\widetilde O_{N,\chi}$ and
$
 \delta_\chi:=\sum_{i>\chi}|a_i|^2
 =D^{-1}\|\Delta O\|_2^2.
$
Top-$\chi$ minimizes the Hilbert--Schmidt error among all expansions with at most $\chi$ Pauli terms, including those with reoptimized coefficients, as in best-term nonlinear approximation~\cite{DeVore1998}.
However, the contributions of many small coefficients may accumulate substantially and thereby affect the total error in the expectation value~\cite{RudolphEtAl2025}. Therefore, quantitative guarantees relating the number of retained Pauli terms $\chi$ to the target accuracy need to be established.
Throughout this paper, Pauli truncation refers to this Top-$\chi$ strategy.

In the remainder of this paper, we study when $O_N$ is efficiently approximable  by $\widetilde{O}_{N,\chi}$ through relating the scaling of OSE to a given accuracy $0<\varepsilon<1$.
We first give the definition of efficient approximability by Pauli truncation.

\begin{definition}
\label{def:Pauli_truncation_efficiency}
Let $\{O_N\}_N$ be a discrete family of Hermitian operators on $N$ qubits,
and let $\{f_N\}_N$ be a corresponding family of functionals.
We say that $\{O_N\}_N$ is efficiently approximable by Pauli truncation
with respect to $\{f_N\}_N$ if, for every $N$ and
every $0<\varepsilon<1$,
there exists an integer $\chi =\text{poly}(N,\varepsilon^{-1})$ such that
$| f_N(O_N) - f_N(\widetilde{O}_{N,\chi})| \leq \varepsilon,$ where $\widetilde O_{N,\chi}=\Top_\chi(O_N).$
\end{definition}

Efficient approximability in Definition~1 depends on
the quantities specified by the functionals $\{f_N\}_N$,
such as expectation values, correlations, and
fixed-order OTOCs.
These different accuracy requirements lead to
different sufficient ranges of the R\'enyi index,
allowing us to characterize how OSE scaling governs
the number of Pauli terms needed for each task.

It is clear that the squared absolute values of the Pauli coefficients  form a probability distribution.
Now, we present some results on relating the tails of the probability distribution to their OSE.

\begin{lemma}
\label{lem:tails}
For the decomposition \eqref{eq:decomposition} and $0<\chi<r$,
the following hold:
\begin{enumerate}[label=(\roman*)]
\item For $0<\alpha<1$,
\begin{equation}
\delta_\chi
\leq
\left(\frac{\chi}{1-\alpha}\right)^{\frac{\alpha-1}{\alpha}}
\cdot2^{  \big(  \frac{1-\alpha}{\alpha}
\M^{(\alpha)}(O_N)  \big)},
\label{eq:tail-alpha-less-one}
\end{equation}
\item For $\alpha>1$,
\begin{equation}
\delta_\chi
\geq
1-
\chi^{\frac{\alpha-1}{\alpha}}\cdot
2^{\big ( -\frac{\alpha-1}{\alpha}\M^{(\alpha)}(O_N) \big )},
\label{eq:tail-alpha-greater-one}
\end{equation}
\item For $\alpha>0$,
\begin{equation}
|a_{\chi+1}|
\leq
(\chi+1)^{-\frac{1}{2\alpha}}\cdot
2^{ \big(  \frac{1-\alpha}{2\alpha}\M^{(\alpha)}(O_N)
 \big) },
\label{eq:first-omitted}
\end{equation}
\item For $0<\alpha<1/2$,
\begin{equation}
\sum_{i>\chi}|a_i|
\leq
\chi^{-\frac{1-2\alpha}{2\alpha}}\cdot
\frac{2\alpha}{1-2\alpha}\,
2^{\frac{1-\alpha}{2\alpha}\M^{(\alpha)}(O_N)}
.
\label{eq:l1-tail}
\end{equation}
\end{enumerate}
\end{lemma}

\begin{proof}
The proofs of parts (i) and (ii) are similar to those in Refs.~\cite{Verstraete2006Matrix,Schuch2008Entropy} and are omitted.
For part (iii), since $\{a_i^2 \}_i$ is a nonincreasing sequence, it follows that
\[j |a^2_j|^\alpha \leq \sum_{i=1}^j |a^2_i|^\alpha \leq  \sum_{i=1}^\infty |a^2_i|^\alpha. \]
Hence
\[|a_j| \leq (\sum_{i=1}^\infty |a^2_i|^\alpha)^ \frac{1}{2\alpha}j^\frac{-1}{2\alpha}=2^{ \big\{  \frac{1-\alpha}{2\alpha}\M^{(\alpha)}(O_N)
-\frac{\log(j) }{2\alpha} \big\} }.
\]
Setting $j=\chi+1$, we have
\[|a_{\chi+1}|
\leq (\chi+1)^{-\frac{1}{2\alpha}}\cdot
2^{ \big(  \frac{1-\alpha}{2\alpha}\M^{(\alpha)}(O_N)
 \big) }.
\]
For part (iv), part (iii) gives
\[
|a_i| \leq 2^{ \big\{  \frac{1-\alpha}{2\alpha}\M^{(\alpha)}(O_N)\big\} }\cdot i^\frac{-1}{2\alpha}.
\]
Therefore, this gives
\[
\sum_{i>\chi}|a_i|
\leq 2^{ \big\{  \frac{1-\alpha}{2\alpha}\M^{(\alpha)}(O_N)\big\} }
\sum_{i=\chi+1}^{\infty}i^{ \frac{-1}{2\alpha}}.
\]
When $0<\alpha<1/2$, the exponent $1/(2\alpha)>1$.  The integral test therefore gives
\[
\sum_{i=\chi+1}^{\infty}i^{ \frac{-1}{2\alpha}}
\leq
\int_\chi^\infty x^{ \frac{-1}{2\alpha}}\,dx\\
=\frac{2\alpha}{1-2\alpha}
\chi^{-\frac{1-2\alpha}{2\alpha}}.
\]
Hence,
\[
\sum_{i>\chi}|a_i|
\leq
\chi^{-\frac{1-2\alpha}{2\alpha}}\cdot
\frac{2\alpha}{1-2\alpha}\,
2^{\frac{1-\alpha}{2\alpha}\M^{(\alpha)}(O_N)}.
\]
This completes the proof.
\end{proof}

\section{Pauli truncation guarantees}
\label{sec:truncation-guarantees}
For the problem of computing Eq.~\eqref{eq:expectation-value}, the spectral norm distance between $O_N$ and $ \widetilde{O}_{N,\chi}$ bounds the error in computing expectation values for any initial state $\rho$
\begin{equation}
|\tr[(O_N - \widetilde{O}_{N,\chi})\rho]| \leq \|\rho\|_1 \|\Delta O\|_\infty = \|\Delta O\|_\infty.
\label{eq:expectation_error}
\end{equation}
where the first inequality follows from H\"older's inequality,
and the last equality follows from the fact
that $\rho$ is positive semidefinite with unit trace.

We first give a lower bound on the number of Pauli terms required to approximate an operator to a specified accuracy for all states, expressed in terms of the OSE.

\begin{theorem}
\label{thm:necessary}
 Let $O_N$ be a Hermitian operator and $\tilde{O}_{N,\chi}$ its Pauli truncation approximation.
Then there exists a quantum state $\rho$, such that the error in its expectation value $\varepsilon := | \operatorname{tr}[(O_N - \tilde{O}_{N,\chi})\rho]|$, the following hold:
\begin{enumerate}[label=(\roman*)]
\item For $\alpha>1$,
\begin{equation*}
\log(\chi)
\geq
\M^{(\alpha)}(O_N)
+\frac{\alpha}{\alpha-1}\log(1-\varepsilon^2),
\end{equation*}
\item For $\alpha=1$,
\begin{equation*}
\log(\chi)\geq \M^{(1)}(O_N)-2\varepsilon^2 N-1.
\label{eq:necessary-shannon-sharp}
\end{equation*}
\end{enumerate}
\end{theorem}

\begin{proof}
Since $O_N$ is a Hermitian operator, the operators $\tilde{O}_{N,\chi}$ and $\Delta O$ are Hermitian.
We now take $\rho := |\psi_{max} \rangle \langle \psi_{max}|$, which is an eigenstate of $\Delta O$ corresponding to an eigenvalue of largest absolute value $\lambda_{max}$.
In this case, the inequality in Eq.~\eqref{eq:expectation_error}  becomes an equality: $|\tr[\Delta O \rho]| = \|\Delta O\|_{\infty}.$

From the standard relations between Schatten $p$-norms, we have
\begin{align}
\|\Delta O\|_\infty
&\geq D^{-\frac12}\|\Delta O\|_2
=D^{-\frac12}\left\|\sum_{i>\chi}^r a_iP_i\right\|_2  \notag\\
&=D^{-\frac12}\sqrt{\sum_{i,j>\chi}^r a_ia_j\tr[P_iP_j]}
=\sqrt{\sum_{i>\chi}^r a_i^2},
\label{eq:tail}
\end{align}
where the last equality follows from $\tr[P_i P_j]=D \delta_{ij}.$

For $\alpha>1$, from Lemma~\ref{lem:tails} (ii) and Eq.~\eqref{eq:tail}, we can obtain
\begin{align}
\varepsilon^2&=\|\Delta O\|_\infty^2\geq\sum_{i>\chi}^r a_i^2=\delta_\chi \notag\\
&\geq1-\chi^{\frac{\alpha-1}{\alpha}}
2^{-\frac{\alpha-1}{\alpha}\M^{(\alpha)}(O_N)}.
\label{eq:varepsilon-delta}
\end{align}
Since $\alpha/(\alpha-1)>0$ and $0 < \varepsilon < 1$, after a straightforward calculation, we arrive at
\[\log\chi
\geq
\M^{(\alpha)}(O_N)
+\frac{\alpha}{\alpha-1}\log(1-\varepsilon^2).
\]

For $\alpha=1$, since $\sum_{i=1}^{r}a_i^2=1,$  $\M^{(1)}(O_N)=-\sum_{i=1}^{r}a_i^2\log a_i^2$ is the Shannon entropy of the probability distribution $\{a_i^2\}_{i=1}^r$.
We define
$q_i:=\frac{a_i^2}{1- \delta_\chi}$ with $i=1,\cdots,\chi$, and
$s_i:=\frac{a_i^2}{\delta_\chi}$ with $i=\chi+1,\cdots,r$.
It follows immediately that $\sum_{i=1}^\chi q_i= \frac{\sum_{i=1}^\chi a_i^2}{1- \delta_\chi}=1$ and
$\sum_{i=\chi+1}^r s_i=\frac{\sum_{i=\chi+1}^r a_i^2}{\delta_\chi}=1$.
Hence, both $\{q_i \}_{i=1}^\chi$ and $\{ s_i\}_{i=\chi+1}^r$ form probability distributions.
The original probability distribution $\{a_i^2\}_{i=1}^r$ is rewritten as
\begin{equation*}
a^2_i =
\begin{cases}
(1 - \delta_\chi) q_i, & i \leq \chi, \\
\delta_\chi s_i, & i > \chi.
\end{cases}
\end{equation*}
The Shannon OSE $\M^{(1)}$ is divided into two parts:
\begin{align*}
\M^{(1)}(O_N)
&=-\sum_{i=1}^{\chi}a_i^2\log a_i^2
  -\sum_{i=\chi+1}^{r}a_i^2\log a_i^2\\
&=-\sum_{i=1}^{\chi}(1-\delta_\chi)q_i
       \log\big[(1-\delta_\chi)q_i\big]\\
&\quad-\sum_{i=\chi+1}^{r}\delta_\chi s_i
       \log(\delta_\chi s_i)\\
&=-(1-\delta_\chi)\sum_{i=1}^{\chi}q_i\log(1-\delta_\chi)\\
&\quad-(1-\delta_\chi)\sum_{i=1}^{\chi}q_i\log q_i\\
&\quad-\delta_\chi\sum_{i=\chi+1}^{r}s_i\log\delta_\chi
       -\delta_\chi\sum_{i=\chi+1}^{r}s_i\log s_i\\
&=-(1-\delta_\chi)\log(1-\delta_\chi)
       +(1-\delta_\chi)H(q)\\
&\quad-\delta_\chi\log\delta_\chi+\delta_\chi H(s)\\
&=h(\delta_\chi)+(1-\delta_\chi)H(q)+\delta_\chi H(s),
\end{align*}
where $H(x)=-\sum_{i=1} x_i \log x_i$ is the Shannon entropy and
$h(x)=-x\log x-(1-x)\log (1-x)$ is the binary entropy.
Since $H(q)\leq \log \chi,$ $H(s)\leq \log (r-\chi)\leq \log (4^N)=2N$, $h(\delta_\chi)\leq 1,$ and $0\leq 1 - \delta_\chi\leq 1$, we have
\begin{equation}
\label{eq:Shannon-entropy}
\M^{(1)}(O_N)\leq 1+\log \chi + 2N \delta_\chi.
\end{equation}
Together with Eq.~\eqref{eq:varepsilon-delta}, this implies
\[\log(\chi)\geq \M^{(1)}(O_N)-2\varepsilon^2 N-1.\]
This completes the proof.

\end{proof}

Now, we interpret Theorem~\ref{thm:necessary} in terms of efficient Pauli truncation as defined in Definition~\ref{def:Pauli_truncation_efficiency}.
Assume  that $\M^{(1)}(O_N)\geq cN$ with some $c>0$.
We choose a fixed error $0<\varepsilon<1$ satisfying $2\varepsilon^2<c$.
Then $\chi\geq 2^{(c-2\varepsilon^2)N-1}$.
Hence $\chi$ must grow exponentially with $N$.
For a fixed $\alpha>1$, we assume  that $\M^{(\alpha)}(O_N)\geq cN^a$ with some $c> 0$ and $a> 0$, then
$\chi\geq
(1-\varepsilon^2)^{\frac{\alpha}{\alpha-1}}2^{cN^a}.$
For every fixed $\varepsilon$, this lower bound grows faster
than any polynomial in $N$.
Therefore, if either $\M^{(1)}(O_N)=\Omega(N)$ or
$\M^{(\alpha)}(O_N)=\Omega(N^a)$ for some fixed $\alpha>1$
and $a>0$, a suitably chosen fixed accuracy requires
superpolynomially many retained Pauli terms $\chi$.
By Definition~\ref{def:Pauli_truncation_efficiency}, the operator
family cannot be efficiently approximated by Pauli truncation
with expectation value accuracy for all states.
For $\alpha>1$, the required lower bound on $\chi$ follows
by combining the squared-coefficient tail lower bound
in Theorem~2 of Ref.~\cite{ShaoChengLiu2026}
with Eq.~\eqref{eq:tail}.
For $\alpha=1$, our result strengthens the lower bound in Eq.~(A32)
of Ref.~\cite{DowlingKosTurkeshi2025}.
For example, if $\M^{(1)}(O_N)=N$ and $\varepsilon=1/2$,
the bound in Ref.~\cite{DowlingKosTurkeshi2025} gives only $\chi\geq1/2$, whereas Theorem~\ref{thm:necessary} yields $\chi\geq2^{N/2-1}$.
Thus, under the same error requirement, our result establishes the necessity of retaining exponentially many Pauli terms, while the bound in
Ref.~\cite{DowlingKosTurkeshi2025} no longer supports this conclusion.

We now consider complementary sufficient conditions.
The Pauli truncation minimizes the Hilbert-Schmidt norm error
(Eq.~\eqref{eq:tail}) which in turn can be related to the OSE (Lemma~\ref{lem:tails}~(i-ii)), whereas the spectral norm provides a universal upper bound for arbitrary expectation values (Eq.~\eqref{eq:expectation_error}).
However, the general relation between the spectral norm and the Hilbert--Schmidt norm is
$D^{-\frac{1}{2}} \|\Delta O\|_2 \leq \|\Delta O\|_\infty \leq \|\Delta O\|_2.$
We therefore control the spectral norm error through
the absolute coefficient tail $\sum_{i>\chi}|a_i|$.

\begin{theorem}
\label{thm:all-state}
Let $O_N$ be a Hermitian operator and $\tilde{O}_{N,\chi}$ its Pauli truncation approximation.
For $0<\alpha<1/2$ and any quantum state $\rho$, the following holds:
\begin{equation*}
\left|\tr[(O_N-\tilde{O}_{N,\chi})\rho]\right|\leq
\frac{2\alpha}{1-2\alpha}\,
\chi^{-\frac{1-2\alpha}{2\alpha}}\cdot2^{\frac{1-\alpha}{2\alpha}\M^{(\alpha)}(O_N)}.
\end{equation*}

\end{theorem}

\begin{proof}
It suffices to obtain an upper bound on $\|O_N-\tilde{O}_{N,\chi}\|_\infty$ by Eq.~\eqref{eq:expectation_error}.
Since every Pauli string is unitary, $\|P_i \|_\infty=1$ for any $P_i\in \mathcal{P}_N$.
Therefore, we have
\begin{equation*}
\begin{aligned}
\|O_N-\tilde O_{N,\chi}\|_\infty
&=\|\Delta O\|_\infty
=\left\|\sum_{i>\chi}a_iP_i\right\|_\infty\\
&\leq\sum_{i>\chi}|a_i|\,\|P_i\|_\infty
=\sum_{i>\chi}|a_i|.
\end{aligned}
\end{equation*}
For $0<\alpha<1/2$, from Lemma~\ref{lem:tails} (iv) and Eq.~\eqref{eq:expectation_error}, we can obtain
\begin{equation*}
\left|\tr[(O_N-\tilde O_{N,\chi})\rho]\right|
\leq\frac{2\alpha}{1-2\alpha}\,
\chi^{-\frac{1-2\alpha}{2\alpha}}
2^{\frac{1-\alpha}{2\alpha}\M^{(\alpha)}(O_N)}.
\end{equation*}
This completes the proof.
\end{proof}

If $\M^{(\alpha)}(O_N)\leq c\log N$ with some fixed $0<\alpha<1/2$ and $c>0$, Theorem~\ref{thm:all-state} implies that $\left|\tr((O_N-\tilde{O}_{N,\chi})\rho)\right|\leq \varepsilon$  is satisfied when  $\chi \geq \left[ \frac{2\alpha}{(1 - 2\alpha)\varepsilon} \right]^{\frac{2\alpha}{1 - 2\alpha}} \cdot N^{\frac{c(1 - \alpha)}{1 - 2\alpha}}.$
For fixed $\alpha$ and $c$, $\chi$ can be chosen polynomial in both $N$ and $\varepsilon^{-1}$.
From Definition~\ref{def:Pauli_truncation_efficiency}, for $\M^{(\alpha)}(O_N)=\mathcal{O}(\log N)$ with $0<\alpha<1/2$,
we have shown that $O_N$ can be efficiently
approximated  using Pauli truncation according to the expectation value for all quantum states.

The restriction $\alpha<1/2$ is essential.
The following example shows that Pauli truncation with unchanged coefficients may require exponentially many terms to achieve a fixed spectral norm accuracy, even when the OSE is bounded for $\alpha\geq1/2$.

\begin{remark}
\label{remark:endpoint}
For any fixed $\alpha\geq1/2$, the condition $\M^{(\alpha)}(O_N)=\mathcal{O}(1)$ alone does not guarantee a Top-$\chi$ truncation satisfying $\|O_N-\widetilde O_{N,\chi}\|_\infty\leq\varepsilon$ with $\chi=\operatorname{poly}(N).$ 
\end{remark}

The claim is proved by the following example.
Since $|0\rangle\langle0|=\frac{I+Z}{2}$, we have that
\begin{equation*}
|0^N\rangle\langle0^N|
=2^{-N}\prod\limits_{i=1}^N(I+Z_i)
=2^{-N}\sum\limits_{S\subseteq \{1,2,\cdots,N\}}Z_S,
\end{equation*}
where $Z_S$ applies $Z$ to the qubits in $S$ and the identity
to the remaining qubits, with $Z_{\varnothing}=I$.
We let
\begin{equation*}
O_N=I-2|0^N\rangle\langle0^N|=(1-2^{1-N})I-2^{1-N}
\sum_{ S\ne\emptyset }Z_S,
\end{equation*}
and
the square of the Pauli coefficients
$(1-2^{1-N})^2+(2^N-1)2^{2-2N}=1$.
Since $O_N^\dagger=O_N$, and  
$O_N^\dagger O_N=O_N^2=I,$ 
$O_N$ is a Hermitian unitary.
Moreover, $\|O_N\|_\infty=1$ and $D^{-1}\tr(O_N^\dagger O_N)=1$,
hence $O_N$ is Hilbert-Schmidt normalized and bounded.

For $N\geq3$, the identity coefficient is positive, and
\begin{equation*}
\begin{aligned}
\M^{ (1/2)}(O_N)
&=2 \log \big[(1-2^{1-N})+(2^N-1)(2^{1-N})\big]\\
&=2 \log (3-2^{2-N})<2 \log (3).
\end{aligned}
\end{equation*}
 Since R\'enyi entropy is nonincreasing in its index, $0\leq\M^{(\alpha)}(O_N)\leq\M^{(1/2)}(O_N)<2\log3,$ for any $\alpha\geq1/2.$
 Hence, the OSE remains bounded independently of $N$ for every
fixed $\alpha\geq1/2$.
In addition, for $N\geq3$, we have $1-2^{1-N}>2^{1-N}$, so the identity
coefficient is uniquely largest in absolute value.
For $1\leq\chi\leq2^N$, Top-$\chi$ truncation therefore retains
the identity and $\chi-1$ nonidentity strings, leaving
$2^N-\chi$ discarded strings.
Since
every $Z_S$ has eigenvalue $+1$ on
$|0^N\rangle$, we have that
\begin{equation*}
\begin{aligned}
\|O_N-\widetilde O_{N,\chi}\|_\infty
&\geq
\left|\langle0^N|(O_N-\widetilde O_{N,\chi})|0^N\rangle\right|\\
&=2^{1-N}(2^N-\chi)
=2\left(1-\frac{\chi}{2^N}\right).
\end{aligned}
\end{equation*}
Conversely, since each Pauli string has spectral norm one,
the triangle inequality gives
$\|O_N-\widetilde O_{N,\chi}\|_\infty
\leq2^{1-N}(2^N-\chi).$
Combining these bounds yields
$\|O_N-\widetilde O_{N,\chi}\|_\infty
=2\left(1-\frac{\chi}{2^N}\right).$
Therefore, achieving any fixed error $0<\varepsilon<1$
requires
$
\chi\geq(1-\frac{\varepsilon}{2})2^N,
$
which grows exponentially with $N$ despite the bounded OSE.
This completes the proof of this claim.

\begin{figure}[htbp]
\centering
\includegraphics[width=\linewidth]{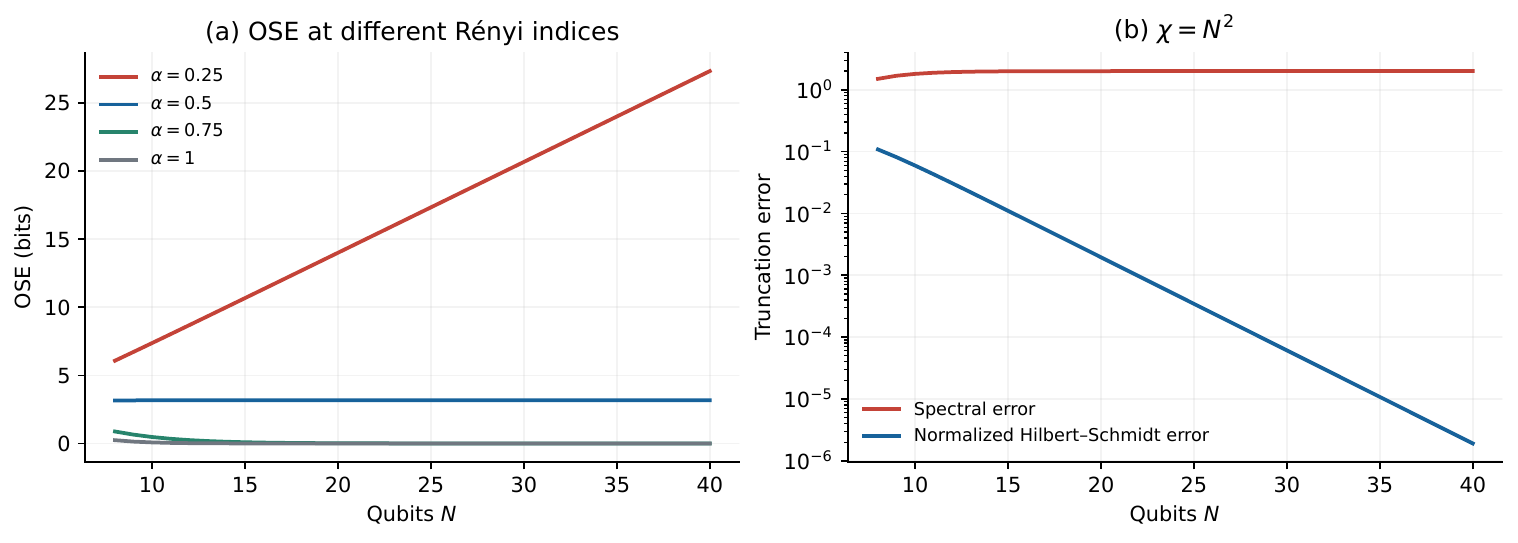}
\caption{OSE and truncation errors for  $O_N=I-2|0^N\rangle\langle0^N|$.
 (a) OSE grows with $N$ for $\alpha<1/2$, remains bounded at $\alpha=1/2$, and tends to zero for $\alpha>1/2$.
 (b) With $\chi=N^2$ retained Pauli terms, the normalized Hilbert--Schmidt error tends to zero, while the spectral norm error approaches two.}
\label{fig:endpoint}
\end{figure}

This example also illustrates the difference between the normalized
Hilbert--Schmidt and spectral norm truncation errors.
For $1<\chi<2^N$, the discarded Pauli weight is
$
\delta_\chi
=
(2^N-\chi)2^{2-2N}
=
2^{2-N}\left(1-\frac{\chi}{2^N}\right).
$
By Pauli orthogonality,
$
D^{-\frac12}\|O_N-\widetilde O_{N,\chi}\|_2
=
\sqrt{\delta_\chi}
=
2^{1-\frac{N}{2}}\sqrt{1-\frac{\chi}{2^N}}.
$
However, all discarded Pauli strings have eigenvalue $+1$ on
$\lvert0^N\rangle$ and carry the same negative coefficient.
Their contributions therefore add without cancellation on this state,
saturating the triangle-inequality bound and giving
$
\|O_N-\widetilde O_{N,\chi}\|_\infty
=
2\left(1-\frac{\chi}{2^N}\right).
$
Consequently, whenever $\chi=\operatorname{poly}(N)$, the normalized
Hilbert--Schmidt error vanishes exponentially with \(N\), whereas the
spectral norm error approaches $2$. The discrepancy arises because
the normalized Hilbert--Schmidt error measures the squared magnitudes
of the discarded coefficients, while the spectral norm detects their
coherent accumulation on a common eigenstate. Thus, a small normalized
Hilbert--Schmidt truncation error does not by itself guarantee accurate
expectation values uniformly over all states.
Figure~\ref{fig:endpoint} illustrates the OSE scaling and the separation between Hilbert--Schmidt and spectral norm truncation errors for the family of operators in Remark~\ref{remark:endpoint}.

Next, we consider restricting the quantum states to a low-average ensemble, which extends the admissible range of the R\'enyi index to $0<\alpha<1$.
First we give the definition of a low-average state ensemble, which is introduced in Refs.~\cite{Dowling2026LOE,Schuster2025Polynomial}.

\begin{definition}
\label{def:low-average-ensemble}
Let $\mathcal E_b$ be an ensemble of quantum states. If, for some $b\geq1$, its first moment satisfies
$\| \E_{\rho\sim\mathcal E_b}\rho \|_\infty\leq \frac{b}{D}$, then
we say that $\mathcal E_b$  is $b$-low-average.
\end{definition}
Examples with $b=1$ include the uniform ensemble of
computational basis states and the singleton ensemble
containing the infinite-temperature state
$\frac{I}{D}$~\cite{Dowling2026LOE}.
\begin{theorem}
\label{thm:low-average}
Let $O_N$ be a Hermitian operator and $\tilde{O}_{N,\chi}$ its Pauli truncation approximation, and
let $\mathcal E_b$ be a $b$-low-average ensemble of quantum states.
For $0<\alpha<1$ and any operator $A$ satisfying $\|A\|_\infty\leq1$, the following holds:
\begin{equation*}
\E_{\rho\sim\mathcal E_b}
\left|\tr((O_N-\tilde{O}_{N,\chi}) A\rho)\right|\\
\leq \sqrt b\,
2^{\frac{1-\alpha}{2\alpha}
\left[\M^{(\alpha)}(O_N)-\log \!\left(\frac{\chi}{1-\alpha}\right)\right]}.
\label{eq:low-average-ose}
\end{equation*}
\end{theorem}
\begin{proof}
We consider the spectral decomposition of quantum state $\rho=\sum_jr_j|j\rangle\langle j|$, $r_j\geq 0$, then $\sqrt \rho=\sum_j \sqrt{ r_j}|j\rangle\langle j|$.
Therefore, it follows that
\begin{align*}
&\E_{\rho\sim\mathcal E_b}|\tr(\Delta OA\rho)|^2\\
&=\E_{\rho\sim\mathcal E_b}|\tr[\Delta OA\sqrt\rho\sqrt\rho]|^2\\
&\leq\E_{\rho\sim\mathcal E_b}
 \Big\{\tr\big[\Delta OA\sqrt\rho(\Delta OA\sqrt\rho)^\dagger\big]
\tr\big[(\sqrt\rho)^\dagger\sqrt\rho\big]\Big\}\\
&=\E_{\rho\sim\mathcal E_b}\tr\big[A^\dagger(\Delta O)^\dagger\Delta OA\rho\big]\\
&=\tr\left[\big(A^\dagger(\Delta O)^\dagger\Delta OA\big)
 \E_{\rho\sim\mathcal E_b}\rho\right]\\
&\leq\left\|\E_{\rho\sim\mathcal E_b}\rho\right\|_\infty
 \big\|A^\dagger(\Delta O)^\dagger\Delta OA\big\|_1\\
&\leq\frac bD\big\|A^\dagger(\Delta O)^\dagger\Delta OA\big\|_1\\
&=\frac bD\|\Delta OA\|_2^2\\
&\leq\frac bD\|\Delta O\|_2^2\|A\|_\infty^2\\
&\leq\frac bD\|\Delta O\|_2^2,
\end{align*}
where the first inequality follows from the Cauchy--Schwarz inequality,
the second inequality follows from H\"older's inequality,
the third inequality follows from the definition of the $b$-low-average ensemble in Definition~\ref{def:low-average-ensemble},
the fourth equality follows from the fact that $\left\| A^\dagger (\Delta O)^\dagger \Delta O A \right\|_1
= \tr \left( A^\dagger (\Delta O)^\dagger \Delta O A \right)= \| \Delta O A \|_2^2,$
since $A^\dagger (\Delta O)^\dagger \Delta O A $ is positive semidefinite.
The fifth inequality  follows from the fact that $\|XY\|_2 \le \|X\|_2 \|Y\|_\infty$,
and the last inequality follows from the boundedness of the  operator $A$.
A similar relation
has been derived in
Refs.~\cite{Dowling2026LOE,Schuster2025Polynomial}.

Finally, from Jensen's inequality and Lemma~\ref{lem:tails} (i), for $0<\alpha<1$, we have that
\begin{equation*}
\begin{aligned}
\E_{\rho\sim\mathcal E_b}
\left|\tr((O_N-\tilde{O}_{N,\chi}) A\rho)\right|
&\leq
\sqrt{
\E_{\rho\sim\mathcal E_b}|\tr(\Delta OA\rho)|^2
}\\
&\leq\sqrt b\,
2^{\frac{1-\alpha}{2\alpha}
\left[\M^{(\alpha)}(O_N)-\log \!\left(\frac{\chi}{1-\alpha}\right)\right]}.
\end{aligned}
\end{equation*}
This completes the proof.
\end{proof}

\begin{figure}[!tbp]
\centering
\includegraphics[width=0.65\linewidth]{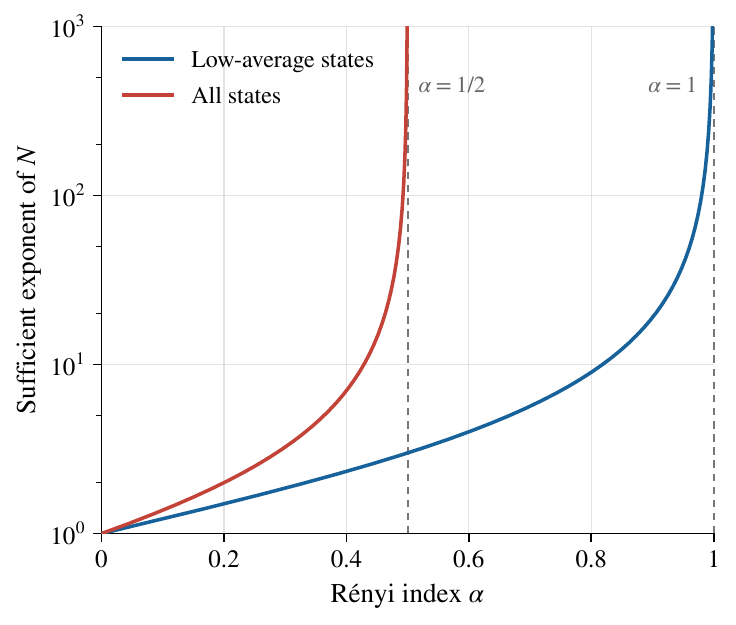}
\caption{Comparison of sufficient Pauli-term $\chi$ exponents for all state and average-state accuracy.
We assume $\M^{(\alpha)}\leq\log N$, $\varepsilon=N^{-1}$, and $b=1$.
The vertical axis shows the power of $N$ in $\chi$: $\frac{1+\alpha}{1-2\alpha}$ for all input
states (red, $0<\alpha<1/2$) and $\frac{1+\alpha}{1-\alpha}$ for low-average ensembles (blue, $0<\alpha<1$).}
\label{fig:budget-exponents}
\end{figure}

If $\M^{(\alpha)}(O_N)\leq c\log N$ with $0<\alpha<1$ and $c>0$, from Theorem~\ref{thm:low-average}, we obtain that
it suffices for ensemble-averaged error at most $\varepsilon$ that
$
\chi\geq
(1-\alpha)N^c
\left(\frac{b}{\varepsilon^2}\right)^{\frac{\alpha}{1-\alpha}}.
$
If $b$ is bounded by a polynomial in $N$ with fixed
$\alpha$ and $c$, this yields that $\chi$ can be chosen polynomial
in both $N$ and $\varepsilon^{-1}$.
From Definition~\ref{def:Pauli_truncation_efficiency}, for $\M^{(\alpha)}(O_N)=\mathcal{O}(\log N)$ with $0<\alpha<1$,
we have shown that $O_N$ can be efficiently
approximated using Pauli truncation on average when $b=\mathcal{O}( \text{poly}(N))$.
Figure~\ref{fig:budget-exponents} compares the sufficient
Pauli-term $\chi$ exponents for accuracy over all states
and average accuracy over low-average ensembles.
The all-state guarantee requires a larger sufficient $\chi$, whose exponent diverges as $\alpha\to1/2$ from below.

\section{Iterative Pauli propagation}
\label{sec:iterative-pauli}

We now extend the spectral norm bound of Theorem~\ref{thm:all-state} to iterative Pauli
propagation and state the additional conditions for efficient readout.

\begin{figure}[!tbp]

\centering
\includegraphics[width=\linewidth]{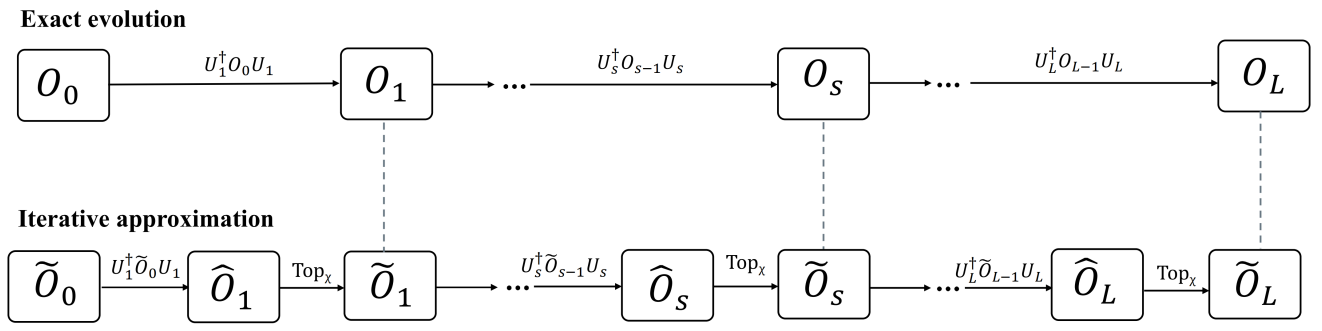}
\caption{Schematic of exact evolution and iterative Pauli propagation.
It starts from $O_0=\widetilde O_0=O$.
The upper trajectory shows the exact operators $O_s$.
Along the lower trajectory, each unitary update produces
the pre-truncation operator
$\widehat O_s=U_s^\dagger\widetilde O_{s-1}U_s$,
followed by Top-$\chi$ truncation to obtain
$\widetilde O_s$.
Dashed lines connect the exact and approximate operators
at the same update step.
}
\label{fig:iterative-schematic}

\end{figure}

Let $U_1,\ldots,U_L$ denote unitary updates in their order of application
in Heisenberg back-propagation. For an integer $\chi\ge1$, we assume that the initial operator
$O$ is Hermitian, satisfies $D^{-1}\tr(O^2)=1$,
and contains at most $\chi$ Pauli terms.
We set
\begin{equation}
\label{eq:iterative-recursion}
\begin{aligned}
O_0&=\widetilde O_0=O,
O_s=U_s^\dagger O_{s-1}U_s,\\
\widehat O_s&=U_s^\dagger\widetilde O_{s-1}U_s,
\widetilde O_s=\Top_\chi(\widehat O_s),\qquad1\leq s\leq L,
\end{aligned}
\end{equation}
 where
 $\widehat O_s$ is the
pre-truncation operator generated from the preceding approximate
operator.
Namely, each step of the approximate evolution consists of a unitary update
followed immediately by Pauli truncation.  At each truncation step, coefficients of identical Pauli strings are first
summed, and only the $\chi$ terms with the largest resulting coefficient
magnitudes are retained.
If fewer than $\chi$ nonzero terms are present, all are retained.
When multiple Pauli strings have coefficients with the same absolute value, ties are broken according to a fixed ordering of Pauli strings.
Figure~\ref{fig:iterative-schematic} illustrates the exact
evolution and the iterative approximation defined above.

To evaluate the OSE of the intermediate operators, we define the
normalization operator
\begin{equation}
\label{eq:normalized-pre-truncation-operator}
A_s=\frac{\widehat O_s}{\nu_s},
\end{equation}
where
$ \nu_s=D^{-\frac{1}{2}}\|\widehat O_s\|_2.$
Hence, $\|A_s\|_2=\|\frac{\widehat O_s}{\nu_s}\|_2=D^{\frac{1}{2}}$, and $\frac{1}{D} \tr(A^\dagger_s A_s)=1$.

\begin{theorem}
\label{thm:iterative-topchi}
For the recursions above, $0<\nu_s\le1$ for every $s$. For any quantum state $\rho$ and
$0<\alpha<1/2$, the following holds:
\begin{equation}
\label{eq:iterative-ose-bound}
\begin{aligned}
\left|\tr((O_L-\widetilde O_L)\rho)\right|
&\leq \sum_{s=1}^L\|\widehat O_s-\widetilde O_s\|_\infty\\
&\leq\frac{2\alpha}{1-2\alpha}
 \chi^{-\frac{1-2\alpha}{2\alpha}}
 \sum_{s=1}^L\nu_s\,
 2^{\frac{1-\alpha}{2\alpha}\mathcal M^{(\alpha)}(A_s)}.
\end{aligned}
\end{equation}
\end{theorem}

\begin{proof}
Subtracting the recursions gives
$
 O_s-\widetilde O_s
 =U_s^\dagger(O_{s-1}-\widetilde O_{s-1})U_s
  +(\widehat O_s-\widetilde O_s).
$
The triangle inequality and unitary invariance of the spectral norm
therefore imply
$
 \|O_s-\widetilde O_s\|_\infty
 \le\|O_{s-1}-\widetilde O_{s-1}\|_\infty
    +\|\widehat O_s-\widetilde O_s\|_\infty.
$
Iterating from $O_0=\widetilde O_0$, we have that
\begin{equation*}
\begin{aligned}
&\left|\tr((O_L-\widetilde O_L)\rho)\right|\\
&\leq\|O_L-\widetilde O_L\|_\infty\\
&\leq\|O_{L-1}-\widetilde O_{L-1}\|_\infty
      +\|\widehat O_L-\widetilde O_L\|_\infty\\
&\leq\|O_{L-2}-\widetilde O_{L-2}\|_\infty\\
&\quad+\|\widehat O_{L-1}-\widetilde O_{L-1}\|_\infty
      +\|\widehat O_L-\widetilde O_L\|_\infty\\
&\leq\cdots\\
&\leq\|O_0-\widetilde O_0\|_\infty
      +\|\widehat O_1-\widetilde O_1\|_\infty\\
&\quad+\cdots+\|\widehat O_L-\widetilde O_L\|_\infty\\
&=\sum_{s=1}^L\|\widehat O_s-\widetilde O_s\|_\infty.
\end{aligned}
\end{equation*}

Since the initial normalized operator $O$  is nonzero, unitary conjugation preserves its nonzero norm. Moreover, each Top-$\chi$ truncation with $\chi\geq1$
retains at least one nonzero coefficient. Thus, all intermediate approximate operators $\widehat O_s$ are nonzero, and consequently $\nu_s>0$.
For each step $s$, we let $\widehat O_s=\sum_j a_{s,j}P_j$ and obtain $\widetilde O_s=\sum_{j\leq\chi} a_{s,j}P_j$.
Since every Pauli string has spectral norm one,
the triangle inequality gives
\begin{equation}
\label{eq:cumulative-error-estimate}
\|\widehat O_s-\widetilde O_s\|_\infty
=\left\|\sum_{j>\chi}a_{s,j}P_j\right\|_\infty
\leq\sum_{j>\chi}|a_{s,j}|.
 \end{equation}
By unitary invariance of the Hilbert--Schmidt norm and Pauli orthogonality, we have
$\|\widetilde O_s\|_2= \sqrt{ D \sum_{j\leq\chi} |a_{s,j}|^2  }\leq \sqrt{ D \sum_{j} |a_{s,j}|^2  }=\|\widehat O_s\|_2=\|U_s^\dagger\widetilde O_{s-1}U_s\|_2=\|\widetilde O_{s-1}\|_2.$
Together with the initial
normalization of $O$, recursively, this gives
\begin{equation}
\label{eq:Hilbert--Schmidt-norm-bound}
0<\nu_s=D^{-\frac12}\|\widetilde O_{s-1}\|_2
\leq D^{-\frac12}\|O\|_2=1,
s=1,\ldots,L.
\end{equation}
By the positive homogeneity of $\operatorname{Top}_{\chi}$,
$\widehat O_s-\widetilde O_s
 =\nu_s\left[
 A_s-\operatorname{Top}_{\chi}(A_s)
 \right].
$
Applying Theorem~\ref{thm:all-state} to each normalized
Hermitian operator $A_s$, and using the positive homogeneity
of $\Top_\chi$, we obtain
\begin{equation*}
\begin{aligned}
\sum_{s=1}^L\|\widehat O_s-\widetilde O_s\|_\infty
&=\sum_{s=1}^L\nu_s\|A_s-\Top_\chi(A_s)\|_\infty\\
&\leq\frac{2\alpha}{1-2\alpha}
 \chi^{-\frac{1-2\alpha}{2\alpha}}
 \sum_{s=1}^L\nu_s\,
 2^{\frac{1-\alpha}{2\alpha}\M^{(\alpha)}(A_s)}.
\end{aligned}
\end{equation*}
 This completes the proof.
\end{proof}

We assume that $\mathcal M^{(\alpha)}(A_s)\le M_*$ throughout the actual
pre-truncation trajectory.
It is therefore sufficient for error at most $\varepsilon>0$ to choose
an integer $\chi$ satisfying
\begin{equation}
 \chi\ge
 \left[
 \frac{2\alpha L}{(1-2\alpha)\varepsilon}
 \right]^{\frac{2\alpha}{1-2\alpha}}\cdot
 2^{\frac{1-\alpha}{1-2\alpha}M_*}.
 \label{eq:iterative-sufficient-budget}
\end{equation}
For fixed $\alpha$, the right-hand side of
Eq.~\eqref{eq:iterative-sufficient-budget} is polynomial in $N$ whenever $L$, $\varepsilon^{-1}$, and
$2^{M_*}$ are polynomial in $N$. In particular, a bound
$M_*=\mathcal{O}(\log N)$ along a polynomial-length trajectory is sufficient
for polynomial-size propagation at inverse-polynomial accuracy.

From Eq.~\eqref{eq:cumulative-error-estimate}, 
we also obtain a computable
upper bound that is no larger than the OSE-based bound:
\begin{equation}
\label{eq:discarded-absolute-sum}
\begin{aligned}
&\left|\tr((O_L-\widetilde O_L)\rho)\right|\\
&\leq \sum_{s=1}^L\sum_{j>\chi}|a_{s,j}|\\
&\leq\frac{2\alpha}{1-2\alpha}
 \chi^{-\frac{1-2\alpha}{2\alpha}}
 \sum_{s=1}^L\nu_s\,
 2^{\frac{1-\alpha}{2\alpha}\mathcal M^{(\alpha)}(A_s)},
\end{aligned}
\end{equation}
where the second inequality follows similarly to the proof of Lemma~\ref{lem:tails}~(iv).

\begin{figure}[!tbp]

\centering
\includegraphics[width=\linewidth]{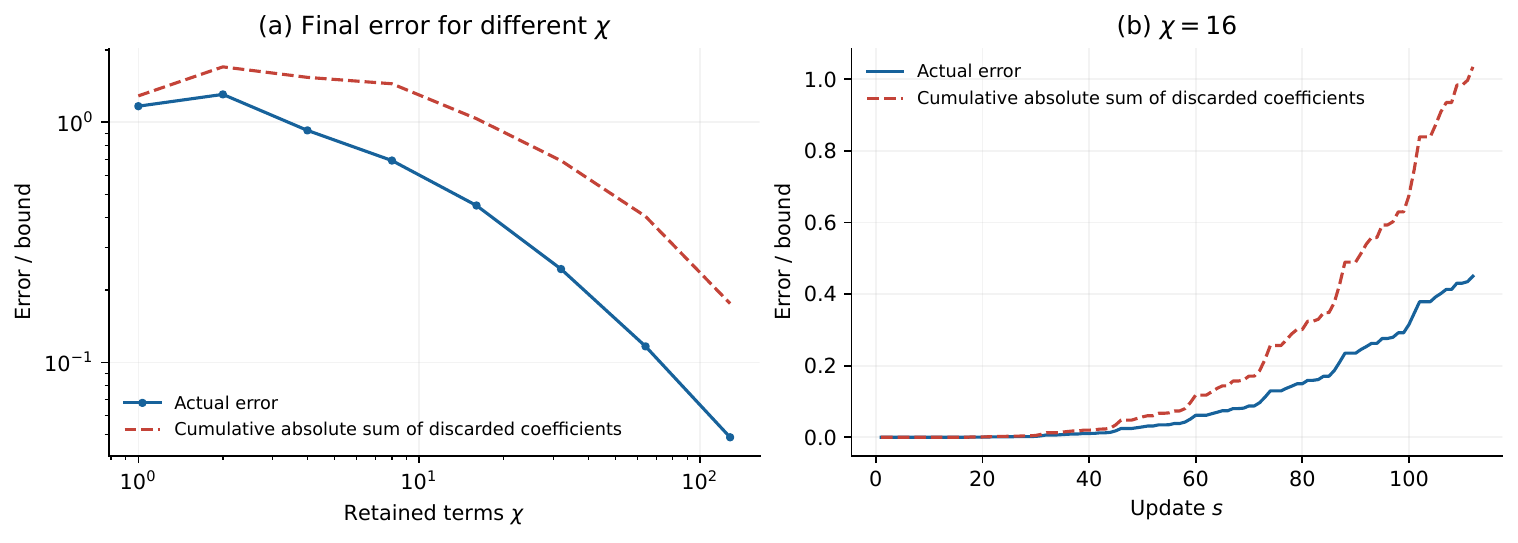}
\caption{Spectral norm error and cumulative absolute sum of discarded
Pauli coefficients
in five-qubit iterative Pauli propagation.
(a) Final error and cumulative bound for different $\chi$.
(b) Their evolution over the update steps for $\chi=16$.
}
\label{fig:iterative}

\end{figure}

Figure~\ref{fig:iterative} compares the spectral norm error
with the cumulative absolute sum of discarded Pauli coefficients in a five-qubit
iterative Pauli propagation example.
Starting from $Z_3$, we repeat eight times the updates
$e^{-\frac{0.17iX_j}{2}}$ for $j=1,\ldots,5$,
$e^{-\frac{0.09iZ_j}{2}}$ for $j=1,\ldots,5$, and
$e^{-\frac{0.13iZ_jZ_{j+1}}{2}}$ for $j=1,\ldots,4$,
in this order, giving 112 updates.
The sequence combines single-qubit $X$ and $Z$
rotations with nearest-neighbor $ZZ$ rotations,
providing a local circuit example for testing
cumulative truncation-error control.
Figure~\ref{fig:iterative}(a) illustrates how the final
approximation error depends on $\chi$,
while Fig.~\ref{fig:iterative}(b) shows how the error
accumulates over successive propagation steps.
Within numerical precision, the spectral norm error does not exceed
the cumulative bound throughout the propagation,
consistent with Theorem~\ref{thm:iterative-topchi}.
This bound can be computed during propagation to assess
the truncation accuracy.

Next, we discuss the computational cost and readout in iterative Pauli propagation.
We assume that each $U_s$ is a Pauli rotation and that the initial
sparse Pauli expansion is explicitly available. Each retained Pauli
term then generates at most two candidates per update. With
hash-based aggregation of identical strings and exact linear-time
Top-$\chi$ selection, the total expected arithmetic cost is
$\mathcal{O}(L\chi N)$, with working storage $\mathcal{O}(\chi N)$.
The final expectation value is evaluated as:
$
 \operatorname{tr}(\rho\widetilde O_L)
 =\sum_{j=1}^{\chi}a_{L,j}\operatorname{tr}(\rho P_j).
$
Efficient readout therefore requires that each retained Pauli
expectation can be computed classically in time polynomial in $N$
and the inverse requested accuracy.
It is clear that product states satisfy this
condition, with $\mathcal{O}(\chi N)$ arithmetic cost for the complete readout.
Moreover, the Hilbert--Schmidt norm bound implied by Eq.~\eqref{eq:Hilbert--Schmidt-norm-bound} gives:
$
 \sum_{j=1}^{\chi}|a_{L,j}|
 \le \sqrt{\chi}\,D^{-\frac{1}{2}}\|\widetilde O_L\|_2
 \le \sqrt{\chi}.
$
Consequently, assigning tolerance $\frac{\varepsilon}{2}$ to propagation
and computing each retained Pauli expectation to additive accuracy
$\frac{\varepsilon}{2\sqrt{\chi}}$ yields total error at most $\varepsilon$.
Together with Eq.~\eqref{eq:iterative-sufficient-budget}, these access
conditions give a classical simulation with polynomial expected
cost in the arithmetic model.

\section{Higher-order correlators}
\label{sec:otoc}
We now consider higher-order out-of-time-ordered correlators (OTOCs) to examine whether Pauli truncation can also accurately approximate higher-order correlations of operator dynamics.
For an integer $k\geq2$,  the
infinite-temperature alternating correlator is defined as
\begin{equation}
\label{eq.OTOC}
\OTOC^{(k)}(O,A):=D^{-1}\tr[(OA)^k],
\end{equation}
 where $O$ is a Hermitian operator and $A$ is a fixed Hermitian operator satisfying $\|A\|_\infty\leq1$.
The case of $k=2$ gives the usual four--point OTOC, which is widely used in the study of information scrambling~\cite{XuSwingle2020,DowlingKosModi2023,Swingle2018}.
Higher-point OTOCs ($k>2$) have been connected to quantum supremacy experiments~\cite{MiEtAl2021,GoogleQuantumAI2025}, unitary designs~\cite{RobertsYoshida2017}, and the foundations of quantum statistical mechanics~\cite{FoiniKurchan2019,Pappalardi2022}.

Throughout this section,  we additionally assume that
$\|O_N\|_\infty\leq1$.
Together with the Hermiticity and normalization imposed
in Section~\ref{sec:setup}, these assumptions imply that
$O_N$ is a Hermitian unitary.

\begin{theorem}
\label{thm:otoc-hs}
Let $O_N$ be a Hermitian unitary operator and $\tilde{O}_{N,\chi}$ its Pauli truncation approximation.
For $0<\alpha<1$, the following holds:
\begin{equation*}
\begin{aligned}
\begin{aligned}
&\left| \OTOC^{(k)}(O_N,A)-\OTOC^{(k)}(\widetilde O_{N,\chi},A) \right|\\
&\leq \Big[ \sum_{j=0}^{k-2} \| \widetilde O_{N,\chi} \|^{j}_\infty +1 \Big]
2^{\frac{1-\alpha}{2\alpha}
\left[\M^{(\alpha)}(O_N)-\log\!\left(\frac{\chi}{1-\alpha}\right)\right]}.
\end{aligned}
\end{aligned}
\end{equation*}

\end{theorem}

\begin{proof}
The proof of this theorem follows a similar approach to that of Theorem~3 in Ref.~\cite{Dowling2026LOE}.
For any matrices $A$ and $B$, by the telescoping sum identity, we have
\begin{equation}
A^{\otimes k}-B^{\otimes k}
=\sum_{j=0}^{k-1}A^{\otimes j}\otimes(A-B)\otimes B^{\otimes(k-1-j)}.
\end{equation}
Let $R$ be a unitary cyclic permutation operator. The following identity holds:
\begin{equation}
\tr\big[(A_1\otimes A_2 \otimes \cdots \otimes A_k)R\big]=\tr\big[A_1 A_2  \cdots  A_k\big].
\end{equation}
We now relate $\left|\OTOC^{(k)}(O_N,A)-\OTOC^{(k)}(\widetilde O_{N,\chi},A)\right|$ to $\|O_N-\widetilde O_{N,\chi}\|_2$.
Applying the above two facts, we can obtain
\begin{align*}
&\left| \OTOC^{(k)}(O_N,A)-\OTOC^{(k)}(\widetilde O_{N,\chi},A) \right|\\
&=\frac1D \left|   \tr\big[(O_NA)^k- (\widetilde O_{N,\chi} A)^k \big ]  \right|\\
&=\frac1D \left|  \tr \big[(O_N^{\otimes k}-\widetilde O_{N,\chi}^{\otimes k})A^{\otimes k} R\big] \right|\\
&=\frac1D \left|  \tr \big[ (\sum_{j=0}^{k-1} O_N ^{\otimes j}\otimes(O_N-\widetilde O_{N,\chi} )\otimes \widetilde O_{N,\chi} ^{\otimes(k-1-j)} )A^{\otimes k} R\big] \right|\\
&=\frac1D\Bigg|\sum_{j=0}^{k-1}\tr\Big[(O_N^{\otimes j}A^{\otimes j})\otimes(O_N-\widetilde O_{N,\chi})A\\
&\hspace{2em}
\otimes\big(\widetilde O_{N,\chi}^{\otimes(k-1-j)}A^{\otimes(k-1-j)}\big)R\Big]\Bigg|\\
&=\frac1D \left|  \sum_{j=0}^{k-1}
\tr\!\left[(O_NA)^j(O_N-\widetilde O_{N,\chi})A
(\widetilde O_{N,\chi}A)^{k-1-j}\right] \right|  \\
&=\frac1D \left|  \sum_{j=0}^{k-1}
\tr\!\left[(O_N-\widetilde O_{N,\chi})A
(\widetilde O_{N,\chi}A)^{k-1-j} (O_NA)^j\right] \right|  \\
&\leq \frac1D  \sum_{j=0}^{k-1}
\|  O_N-\widetilde O_{N,\chi}   \|_2
\|    A(\widetilde O_{N,\chi}A)^{k-1-j} (O_NA)^j  \|_2  \\
&= \frac1D \|  O_N-\widetilde O_{N,\chi}   \|_2
\Big[ \sum_{j=0}^{k-2} \|    A(\widetilde O_{N,\chi}A)^{k-1-j} (O_NA)^j  \|_2 \\
&\hspace{2em} +   \|  A(O_NA)^{k-1}  \|_2  \Big]  \\
&\leq \frac1D \|  O_N-\widetilde O_{N,\chi}   \|_2
\Big[ \sum_{j=0}^{k-2}  \| A \|^k_\infty \| \widetilde O_{N,\chi} \|^{k-2-j}_\infty \| \widetilde O_{N,\chi}\|_2\\
&\hspace{2em} \times \| O_N \|^j_\infty+\| A \|^k_\infty  \| O_N \|^{k-1}_\infty  \| I\|_2 \Big]\\
&\leq \frac{1}{\sqrt{D}} \|  O_N-\widetilde O_{N,\chi}   \|_2
\Big[ \sum_{j=0}^{k-2} \| \widetilde O_{N,\chi} \|^{j}_\infty +1 \Big],
\end{align*}
where the first inequality follows from the Hilbert--Schmidt Cauchy--Schwarz inequality,
the second inequality follows from the inequality  $\|AB\|_2\leq \|A\|_\infty\|B\|_2$,
the third inequality follows from the assumptions $\|A\|_\infty\leq1$ and $\|O_N\|_\infty\leq1$, the inequality $\|  \widetilde O_{N,\chi}   \|_2 \leq \sqrt{D}$
and the equality $\| I\|_2=\sqrt{D}$.

For $0<\alpha<1$, from Lemma~\ref{lem:tails} (i), we have that
\begin{equation*}
\begin{aligned}
&\left| \OTOC^{(k)}(O_N,A)-\OTOC^{(k)}(\widetilde O_{N,\chi},A) \right|\\
&\leq \Big[ \sum_{j=0}^{k-2} \| \widetilde O_{N,\chi} \|^{j}_\infty +1 \Big]
2^{\frac{1-\alpha}{2\alpha}
\left[\M^{(\alpha)}(O_N)-\log\!\left(\frac{\chi}{1-\alpha}\right)\right]}.
\end{aligned}
\end{equation*}
Moreover, when $\| \widetilde O_{N,\chi} \|_\infty=1$,
\begin{equation*}
\begin{aligned}
&\left| \OTOC^{(k)}(O_N,A)-\OTOC^{(k)}(\widetilde O_{N,\chi},A) \right|\\
&\leq k
2^{\frac{1-\alpha}{2\alpha}
\left[\M^{(\alpha)}(O_N)-\log\!\left(\frac{\chi}{1-\alpha}\right)\right]},
\end{aligned}
\end{equation*}
when $\| \widetilde O_{N,\chi} \|_\infty\neq1$,
\begin{equation*}
\begin{aligned}
&\left| \OTOC^{(k)}(O_N,A)-\OTOC^{(k)}(\widetilde O_{N,\chi},A) \right|\\
&\leq \Big( \frac{\| \widetilde O_{N,\chi} \|^{k-1}_\infty -1}{\| \widetilde O_{N,\chi} \|_\infty -1}
 +1 \Big)
2^{\frac{1-\alpha}{2\alpha}
\left[\M^{(\alpha)}(O_N)-\log\!\left(\frac{\chi}{1-\alpha}\right)\right]}.
\end{aligned}
\end{equation*}

This completes the proof.

\end{proof}

For $k=2$, we let $0<\alpha<1$ and $c>0$
be fixed constants independent of $N$.
If $\M^{(\alpha)}(O_N)\leq c\log N$, then
Theorem~\ref{thm:otoc-hs} guarantees a fixed error
at most $\varepsilon$ whenever
$
\chi\geq
(1-\alpha)N^c
\left(\frac{2}{\varepsilon}\right)^{\frac{2\alpha}{1-\alpha}}.
$
Thus, four-point OTOCs are efficiently approximable by Pauli truncation with a number of retained terms polynomial in $N$ and $\varepsilon^{-1}$.

For fixed $k\geq3$ and $0<\alpha<1$, we have
$
1+\sum_{j=0}^{k-2}
\|\widetilde O_{N,\chi}\|_\infty^j
\leq
k\max\{1,\|\widetilde O_{N,\chi}\|_\infty\}^{k-2}.
$
Since the Pauli coefficients satisfy
$\sum_j|a_j|^2=1$, the Cauchy--Schwarz inequality gives
$
\|\widetilde O_{N,\chi}\|_\infty
\leq\sum_{j=1}^{\chi}|a_j|
\leq \sqrt{\chi}\sqrt{ \sum_{j=1}^{\chi}|a_j|^2 }
\leq\sqrt{\chi}.
$
Consequently, Theorem~\ref{thm:otoc-hs} yields
\begin{equation*}
\begin{aligned}
&\left|
\OTOC^{(k)}(O_N,A)
-\OTOC^{(k)}(\widetilde O_{N,\chi},A)
\right|\\
&\leq k(1-\alpha)^{\frac{1-\alpha}{2\alpha}}
2^{\frac{1-\alpha}{2\alpha}\M^{(\alpha)}(O_N)}
\chi^{-\frac{1-(k-1)\alpha}{2\alpha}}.
\end{aligned}
\end{equation*}
If $\M^{(\alpha)}(O_N)\leq c\log N$
with some fixed $c>0$, then for
$0<\alpha<\frac{1}{k-1}$, an error at most
$\varepsilon$ is guaranteed whenever
\[
\chi\geq
\left[
\frac{k(1-\alpha)^{\frac{1-\alpha}{2\alpha}}}
{\varepsilon}
N^{\frac{c(1-\alpha)}{2\alpha}}
\right]^{\frac{2\alpha}{1-(k-1)\alpha}}.
\]
Thus, higher-order OTOCs are efficiently
approximable by Pauli truncation in this range
with a number of retained terms polynomial
in $N$ and $\varepsilon^{-1}$.

However,
for general $0<\alpha<1$, Theorem~\ref{thm:otoc-hs}
can yield a polynomial $\chi$ if
$\|\widetilde O_{N,\chi}\|_\infty$
admits a polynomial upper bound in $N$
that is independent of $\chi$.
We therefore consider an alternative approach
based on the spectral-norm truncation error.

\begin{theorem}
\label{thm:otoc-uniform}

Let $O_N$ be a Hermitian unitary operator and $\tilde{O}_{N,\chi}$ its Pauli truncation approximation.
For $0<\alpha<1/2$, the following holds:
\begin{equation*}
\begin{aligned}
&\left|\OTOC^{(k)}(O_N,A)-\OTOC^{(k)}(\widetilde O_{N,\chi},A)\right|\\
&\leq \left( 1 + \frac{2\alpha}{1 - 2\alpha} \, 2^{\frac{1-\alpha}{2\alpha} \M^{(\alpha)}(O_N)} \, \chi^{-\frac{1-2\alpha}{2\alpha}} \right)^{\!k} - 1.
\label{eq:otoc-uniform}
\end{aligned}
\end{equation*}
\end{theorem}

\begin{proof}
We set $X=O_NA$ and $Y=\widetilde O_{N,\chi}A$.
Using the noncommutative telescoping identity,
$X^k-Y^k=\sum_{j=0}^{k-1}X^j(X-Y)Y^{k-1-j}.$
By the triangle inequality, we have $\| \widetilde O_{N,\chi} \|_\infty= \|O_N-(O_N-\widetilde O_{N,\chi})\|_\infty\leq \|O_N\|_\infty +\|\Delta O \|_\infty \leq 1+ \|\Delta O \|_\infty,$ since $\|O_N\|_\infty\leq1.$
Applying the above facts together with the inequality $D^{-1} |\tr A|\leq \| A \|_\infty$, we can obtain
\begin{equation*}
\begin{aligned}
&\left| \OTOC^{(k)}(O_N,A)-\OTOC^{(k)}(\widetilde O_{N,\chi},A) \right|\\
&=\frac1D \left|   \tr\big[(O_NA)^k- (\widetilde O_{N,\chi} A)^k \big ]  \right|\\
&\leq \| (O_NA)^k- (\widetilde O_{N,\chi} A)^k\|_\infty\\
&= \| \sum_{j=0}^{k-1}(O_NA)^j\big[(O_NA)-(\widetilde O_{N,\chi} A)\big](\widetilde O_{N,\chi} A)^{k-1-j}  \|_\infty\\
&\leq \|O_N-  \widetilde O_{N,\chi} \|_\infty
\sum_{j=0}^{k-1}  \| O_N \|^j_\infty \| \widetilde O_{N,\chi} \|^{k-1-j}_\infty  \|A \|^k_\infty\\
&\leq  \|\Delta O\|_\infty \sum_{j=0}^{k-1} (1+\|\Delta O\|_\infty )^{k-1-j}\\
&=(1+\|\Delta O\|_\infty )^k-1,
\end{aligned}
\end{equation*}
where the second inequality follows from the inequality  $\|AB\|_\infty\leq \|A\|_\infty\|B\|_\infty$,
and the third inequality follows from  the assumptions $\|A\|_\infty\leq1$.
Applying Theorem~\ref{thm:all-state}, we have
\begin{equation*}
\begin{aligned}
&\left|\OTOC^{(k)}(O_N,A)-\OTOC^{(k)}(\widetilde O_{N,\chi},A)\right|\\
&\leq \left( 1 + \frac{2\alpha}{1 - 2\alpha} \, 2^{\frac{1-\alpha}{2\alpha} \M^{(\alpha)}(O_N)} \, \chi^{-\frac{1-2\alpha}{2\alpha}} \right)^{\!k} - 1.
\end{aligned}
\end{equation*}
This completes the proof.

\end{proof}

Although the OSE index is restricted to $0<\alpha<1/2$,
this bound does not require a separate assumption on
$\|\widetilde O_{N,\chi}\|_\infty$.
Let $k\geq2$, $0<\alpha<1/2$, and $c>0$ be fixed
constants independent of $N$, and assume that
$\M^{(\alpha)}(O_N)\leq c\log N$.
For $0<\varepsilon<1$,
Theorem~\ref{thm:otoc-uniform} guarantees
$
\left|
\OTOC^{(k)}(O_N,A)
-\OTOC^{(k)}(\widetilde O_{N,\chi},A)
\right|
\leq\varepsilon
$
whenever
$
\chi\geq
\left[
\frac{2\alpha}
{(1-2\alpha)[(1+\varepsilon)^{1/k}-1]}
\right]^{\frac{2\alpha}{1-2\alpha}}
N^{\frac{c(1-\alpha)}{1-2\alpha}}.
$
For fixed $k$, the reciprocal of
$(1+\varepsilon)^{\frac{1}{k}}-1$ is $\mathcal{O}(\varepsilon^{-1})$.
Thus, if $\M^{(\alpha)}(O_N)=\mathcal O(\log N)$
for a fixed $0<\alpha<1/2$, a number of retained Pauli
terms polynomial in $N$ and $\varepsilon^{-1}$ suffices
to approximate every fixed-order correlator.
This establishes efficient Pauli truncation approximation
of these correlators according to
Definition~\ref{def:Pauli_truncation_efficiency}.

\section{Spectral norm truncation error under random signs}
\label{sec:analysis-error}

We now call back the problem of the spectral norm error of Pauli truncation.
Theorem~\ref{thm:all-state} guarantees that logarithmic OSE growth allows efficient Pauli truncation to achieve the prescribed accuracy for all input states when
$0<\alpha<1/2$, while Remark~\ref{remark:endpoint} shows that this range cannot be extended using OSE scaling alone.
The reason is that OSE depends only on the squared magnitudes of the Pauli coefficients, and in turn controls the normalized
Hilbert--Schmidt truncation error, whereas the spectral norm error
also depends on how the discarded Pauli terms combine.

To investigate the role of cancellation, we introduce an
independent-sign model in which the Pauli strings and
coefficient magnitudes are fixed, while the signs are
independent and uniformly distributed.
This preserves the OSE and the normalized
Hilbert--Schmidt truncation error, allowing us to establish
spectral-norm error bounds in expectation and with high
probability for $0<\alpha<1$.

We fix distinct Pauli strings
$P_1,\ldots,P_{4^N}\in\mathcal P_N$ and coefficient magnitudes $|a_1|\geq\cdots\geq|a_{4^N}|\geq0$ satisfying $\sum_{i=1}^{4^N}|a_i|^2=1$.
Then we define
\[
 O_N=\sum_{i=1}^{4^N}|a_i|s_iP_i,
 \qquad
 \widetilde O_{N,\chi}
 =\sum_{i=1}^{\chi}|a_i|s_iP_i,
\]
where the signs $s_i\in\{-1,+1\}$ are independent
and take either value with equal probability,
and $\chi$ is an integer satisfying $1\leq\chi\leq4^N$.
Coefficients of equal magnitude are ordered by a fixed
rule independent of the signs.
Thus, $\widetilde O_{N,\chi}$ is the Top-$\chi$
Pauli truncation of every realization of $O_N$,
with the retained coefficients unchanged.
By Pauli orthogonality, we have
\begin{equation*}
D^{-1/2}\|O_N\|_2=1,
\qquad
D^{-1}\|O_N-\widetilde O_{N,\chi}\|_2^2
=\sum_{i>\chi}|a_i|^2.
\end{equation*}
These identities hold for every realization,
so sign randomization preserves both the
Hilbert--Schmidt normalization and the
normalized Hilbert--Schmidt truncation error.
We restrict attention to the nontrivial case
$\sum_{i>\chi}|a_i|^2>0$.
Otherwise, all discarded coefficients vanish and
$\widetilde O_{N,\chi}=O_N$, so the truncation error is zero.
For $0<\alpha<1$, every realization also has the same OSE,
$
\M^{(\alpha)}(O_N)
=\frac{1}{1-\alpha}
\log\!\left(\sum_{i=1}^{4^N}|a_i|^{2\alpha}\right).
$

We use the following Hermitian specialization of the
matrix Rademacher concentration inequality of Theorem~4.1.1 in Ref.~\cite{Tropp2015MatrixConcentration}.

\begin{lemma}
\label{lem:matrix_rademacher}
Let $\{X_i\}$ be a finite sequence of fixed Hermitian
$D\times D$ matrices, and let $\{s_i\}$ be independent
random signs taking the values $+1$ and $-1$
with equal probability.
Then
\[
\mathbb E_s\left\|\sum_i s_iX_i\right\|_\infty
\leq
\sqrt{
2\ln(2D)
\left\|\sum_i X_i^2\right\|_\infty
}.
\]
If $\left\|\sum_i X_i^2\right\|_\infty>0$,
then, for every $u\geq0$,
\begin{equation*}
\Pr_s\!\left\{
\left\|\sum_i s_iX_i\right\|_\infty\geq u
\right\}
\leq \min\left\{
1,\,
2D\exp\!\left(
\frac{-u^2}{
2\left\|\sum_i X_i^2\right\|_\infty
}
\right)
\right\}.
\end{equation*}
If $\sum_i X_i^2=0$, then $\sum_i s_iX_i=0$
for every realization.

\end{lemma}

\begin{theorem}
For the random-sign ensemble above, and $0<\alpha<1$,
\label{thm:random}
\begin{equation*}
\mathbb E_s
\|O_N-\widetilde O_{N,\chi}\|_\infty
\leq \sqrt{2(N+1)\ln2}\,
2^{\frac{1-\alpha}{2\alpha}
\left[
\M^{(\alpha)}(O_N)
-\log\!\left(\frac{\chi}{1-\alpha}\right)
\right]}.
\end{equation*}
Moreover, for every $u\geq0$,
\begin{equation*}
\begin{aligned}
&\Pr_s\!\left\{
\|O_N-\widetilde O_{N,\chi}\|_\infty\geq u
\right\}\\
&\leq \min\left\{
1,\,
2^{N+1}\exp\!\left(
-\frac{u^2}{2\sum_{i>\chi}|a_i|^2}
\right)
\right\}.
\end{aligned}
\end{equation*}

\end{theorem}

\begin{proof}
We consider the difference $\Delta O = O_N - \tilde{O}_{N,\chi}=\sum_{i> \chi}^{4^N}|a_i|s_i P_i$, and set $X_i=|a_i| P_i$ for $i= \chi+1,\cdots, 4^N$.
These are fixed Hermitian matrices, and the signs $s_i$ are independent and uniformly distributed
over $\{-1,+1\}$.
Since $P_i P_i^\dagger=I$, we have
\begin{equation*}
\left\|\sum_{i>\chi} X_i^2\right\|_\infty=\left\| \sum_{i>\chi} (|a_i| P_i)(|a_i| P_i)^\dagger\right\|_\infty= \sum_{i>\chi}  |a_i|^2.
\end{equation*}
By Lemmas~\ref{lem:tails} and~\ref{lem:matrix_rademacher}, for $0<\alpha<1$, this yields
\begin{equation*}
\begin{aligned}
\mathbb{E}_s\|\Delta O\|_\infty
&\leq
\sqrt{
2\ln(2D)
\left\|\sum_{i>\chi}X_i^2\right\|_\infty
}\\
&=
\sqrt{
2\ln(2D)\sum_{i>\chi}|a_i|^2
}\\
&=
\sqrt{2(N+1)\ln2}\,
\left(\sum_{i>\chi}|a_i|^2\right)^{1/2}\\
&\leq
\sqrt{2(N+1)\ln2}\,
\left(\frac{1-\alpha}{\chi}\right)^{
\frac{1-\alpha}{2\alpha}}
2^{\frac{1-\alpha}{2\alpha}\M^{(\alpha)}(O_N)}\\
&=
\sqrt{2(N+1)\ln2}\,
\Big(2^{\frac{1-\alpha}{2\alpha}
\left[
\M^{(\alpha)}(O_N)
-\log\!\left(\frac{\chi}{1-\alpha}\right)
\right]}\Big).
\end{aligned}
\end{equation*}
Here we used $D=2^N$ and
$\ln(2D)=(N+1)\ln2$.
The probability bound follows from the tail inequality
in Lemma~\ref{lem:matrix_rademacher}.
This completes the proof.
\end{proof}

The advantage of this model is that Pauli orthogonality remains exact after sign randomization, and hence both the OSE of $O_N$
and $\sum_{i>\chi}  |a_i|^2$ remain unchanged.
Together  with Eq.~\eqref{eq:expectation_error},
the result  controls the worst-state expectation value
error averaged over the independent signs.
Indeed, if $\M^{(\alpha)}(O_N)\leq c\log N$ with some fixed $0<\alpha<1$ and $c>0$, then
$\mathbb E_s
\|O_N-\widetilde O_{N,\chi}\|_\infty
\leq
\sqrt{2(N+1)\ln2}\,
\left(\frac{(1-\alpha)N^c}{\chi}\right)^{
\frac{1-\alpha}{2\alpha}}.$
Consequently, an average spectral norm error of at most $\varepsilon$
is guaranteed whenever
$  \chi\geq
(1-\alpha)N^c
\left[
\frac{2(N+1)\ln2}{\varepsilon^2}
\right]^{\frac{\alpha}{1-\alpha}}.$
Thus, a number of retained Pauli terms polynomial in $N$ and $\varepsilon^{-1}$ suffices.
The averaging is over the random signs, while the spectral norm controls the expectation value error over all states.
The probability bound in Theorem~\ref{thm:random} also provides a high-probability guarantee simultaneously for all input states.
Specifically,
given a target error $u$ and an allowed failure probability $\eta$, applying Lemma~\ref{lem:tails}(i) and Theorem~\ref{thm:random},
it suffices to choose
$
\chi\geq
(1-\alpha)2^{\mathcal M^{(\alpha)}(O_N)}
\left[
\frac{2\ln(2D/\eta)}{u^2}
\right]^{\frac{\alpha}{1-\alpha}}
$
to ensure
$
\Pr_s\!\left\{
\|O_N-\widetilde O_{N,\chi}\|_\infty
\leq u
\right\}\geq1-\eta.
$

\begin{figure}[!tbp]
\centering
\includegraphics[width=\linewidth]{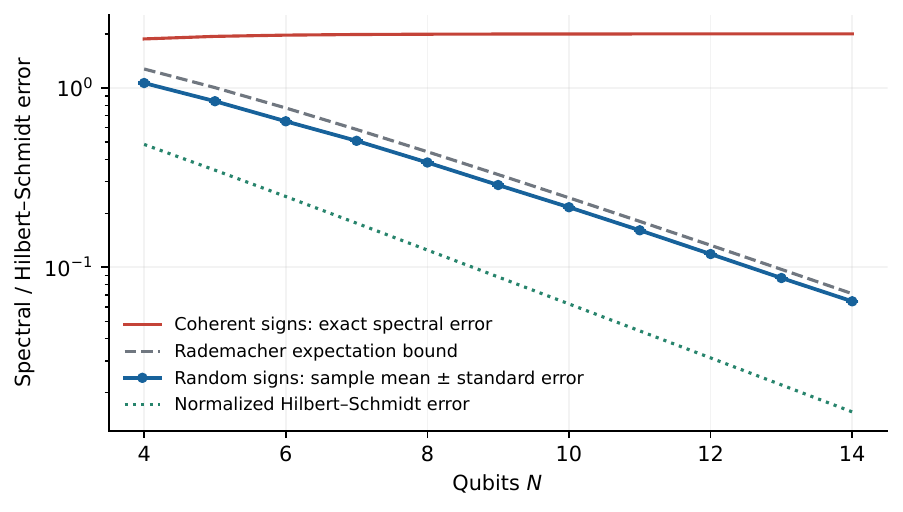}
\caption{Effect of coefficient signs on the spectral-norm
truncation error.
We use the coefficient magnitudes of
$O_N=I-2|0^N\rangle\langle0^N|$ and retain only
the identity term ($\chi=1$).
For each $N$, the random-sign curve shows the sample mean
over 256 independent sign realizations.
Error bars indicate one standard error of the mean,
computed as the sample standard deviation divided by
the square root of the sample size.
The dashed curve is the Rademacher upper bound on
the expected spectral-norm error,
$\sqrt{2(N+1)\ln2}\sqrt{\delta_\chi}$.
}
\label{fig:random-signs}
\end{figure}

Figure~\ref{fig:random-signs} illustrates the effect
of coefficient signs using the family $O_N=I-2|0^N\rangle\langle0^N|$ in Remark~\ref{remark:endpoint}, with only the identity
term retained ($\chi=1$).
The coherent and randomized sign choices have identical
coefficient magnitudes, OSE, and normalized
Hilbert--Schmidt truncation error.
Nevertheless, the coherent-sign spectral error
approaches $2$, whereas the mean spectral error
under independent randomization decreases with $N$.
The figure compares this sampled mean with the
theoretical expectation bound.
This comparison shows that coefficient signs can
substantially affect the spectral norm truncation error,
even when the coefficient magnitudes and OSE are identical,
because cancellation among discarded terms can reduce
the spectral error.

We stress that the independent-sign assumption is
additional statistical information that does not
follow from an OSE bound.
Sign randomization need not preserve unitarity,
so this ensemble does not directly describe
generic local quantum dynamics.
It remains an open question to identify physically
relevant dynamics for which the Pauli truncation error
obeys an analogous spectral norm concentration bound.

\section{Conclusion and discussion}
\label{sec:conclusion}

In this paper, we have established rigorous connections between the scaling of operator stabilizer R\'enyi entropy (OSE) and the efficiency of Pauli truncation.
We derived necessary bounds showing that linear growth of OSE at $\alpha=1$, or power-law growth at $\alpha>1$, rules out efficient Pauli truncation with small expectation value errors over all input states.
Conversely, logarithmic OSE growth for $0<\alpha<1/2$ guarantees polynomial-term Pauli truncation with uniform accuracy.
A counterexample demonstrates that this sufficient range is sharp: even bounded OSE for $\alpha\geq1/2$ does not guarantee efficient approximation.
However, for correlations averaged over low-average state
ensembles whose low-average parameter is polynomially bounded
in the system size, the sufficient range extends to
$0<\alpha<1$.
These results reveal a fundamental distinction between the compressibility of Pauli coefficients and the accuracy of their truncated representation, and clarify how the simulability guarantees depend on the R\'enyi index and the required notion of accuracy.

We further extended the analysis to iterative Pauli propagation and higher-order out-of-time-ordered correlators (OTOCs). For iterative propagation, we established cumulative error bounds that account for successive truncations and provide an explicit way to assess approximation accuracy along the truncated evolution. For fixed-order correlators, we obtained approximation guarantees from OSE scaling, extending the relevance of Pauli truncation beyond expectation values. We also clarified the additional computational requirements needed to translate polynomial truncation budgets into efficient classical simulation algorithms.
Finally, using an independent-sign model, we established spectral norm error bounds in expectation and with high probability for $0<\alpha<1$.
This result demonstrates that cancellation among Pauli terms can substantially reduce the expectation value error over all states, beyond what can be guaranteed by
coefficient magnitudes alone.

Several questions remain.
Since independent sign randomization does not necessarily preserve the structure of physical quantum dynamics, an important direction for future research is to determine whether comparable cancellation properties arise in physically relevant operator evolutions, including random quantum circuits and interacting many-body systems~\cite{AngrisaniEtAl2025,BegusicChan2025}.
Another direction is to establish OSE bounds along truncated trajectories for concrete many-body systems, motivated by recent studies of sparse Pauli dynamics
and operator complexity~\cite{ShaoChengLiu2026}, thereby connecting the present approximation criteria more directly to the practical efficiency of Pauli propagation.

\section*{Acknowledgements}
This paper was supported by National Natural Science Foundation of China (Grant No.~12471437), Postgraduate Education Comprehensive Reform Research and Practice Project of Shaanxi Province in 2025 (Grant No.~YJSZG2025062) and General Projects of Degree and Graduate Education Research in Shaanxi Province in 2026 (Grant No.~SXGERC2026058).

\section*{Data availability}
There are no publicly available research data or software
supporting this manuscript. Requests for further information
or data should be sent to the authors.

\begingroup
\hypersetup{urlcolor=blue}

\endgroup

\end{document}